\documentclass[12pt]{article}

\usepackage{amsmath,epsfig}

\newcommand{\st}{{\tilde{t}}}
\newcommand{\sta}{{\tilde{t}_1}}

\newcommand{\mst}{m_{\tilde{t}_1}}

\newcommand{\cha}{\tilde{\chi}}
\newcommand{\neu}{\tilde{\chi}^0}

\newcommand{\mneu}[1]{m_{\tilde{\chi}^0_{#1}}}

\newcommand{\stopone} {\tilde{t}_1}

\def\mathswitch#1{\relax\ifmmode#1\else$#1$\fi}
\def\mathswitchr#1{\relax\ifmmode{\mathrm{#1}}\else$\mathrm{#1}$\fi}
\newcommand{\PW}{\mathswitchr W}

\newcommand{\MW}{\mathswitch {m_\PW}}

\newcommand{\as}{\alpha_{\rm s}}

\newcommand{\RR}{{\rm R}}

\newcommand{\lesim}{\,\raisebox{-.1ex}{$_{\textstyle <}\atop^{\textstyle\sim}$}\,}

\newcommand{\mycaption}[1]{\caption{\sl #1}}

\begin{document}
\thispagestyle{empty}

\def\thefootnote{\fnsymbol{footnote}}

\begin{flushright}
ZU--TH 14/07
\end{flushright}

\vspace{1cm}

\begin{center}

{\Large\bf Radiative corrections to bino-stop co-annihilation}
       \\[3.5em]
{\large %\sc
A.~Freitas}

\vspace*{1cm}

{\sl
Institut f\"ur Theoretische Physik,
        Universit\"at Z\"urich, \\ Winterthurerstrasse 190, CH-8057
        Z\"urich, Switzerland}

\end{center}

\vspace*{2.5cm}

\begin{abstract}

In co-annihilation scenarios, the weakly interacting dark matter particle (WIMP)
is close in mass to another particle that can decay into the WIMP. As a result,
the other particle does not freeze out before the WIMP in the early universe,
and both contribute to the effective dark matter annihilation cross-section.
Since the heavier particle does not need to be weakly interacting, the
co-annihilation processes are in general subject to sizeable radiative
corrections. Here this is analyzed for the example of neutralino-stop co-annihilation in
supersymmetry. The leading QCD corrections are calculated and it is found that
they have a large effect on the effective annihilation cross-section, reaching
more than 50\% in some regions of parameter space.

\end{abstract}

\def\thefootnote{\arabic{footnote}}
\setcounter{page}{0}
\setcounter{footnote}{0}

\newpage

%%%%%%%%%%%%%%%%%%%%%%%%%%%%%%%%%%%%%%%%%%%%%%%%%%%%%%%%%%%%%%
%%%%%%%%%%%%%%%%%%%%%%%%%%%%%%%%%%%%%%%%%%%%%%%%%%%%%%%%%%%%%%

In recent years, the existence of dark matter in the universe has been firmly
established by various different astrophysical experimental methods. From the
cosmic microwave background and large scale structure, the dark matter density
is determined to be $\Omega_{\rm CDM} h^2 = 0.1106^{+0.0056}_{-0.0075}$
\cite{cos}, where $\Omega_{\rm CDM}$ is the ratio of the dark matter energy
density to the critical density $\rho_c = 2H_0^2/(8\pi G_{\rm N})$ with the
Hubble constant $H_0 = h\times 100$~km/s/Mpc and Newton's constant $G_{\rm N}$.
A promising explanation for the nature of dark matter in agreement with direct
observations and simulations of galaxy formation are weakly interacting massive
particles (WIMPs). While the Standard Model does not encompass an appropriate
particle for this purpose, a suitable candidate can arise from many extensions
of the Standard Model, such as supersymmetry, extended gauge groups or
non-anomalous global symmetry groups.

Due to the requirement that the dark matter particle is weakly interacting, its
annihilation cross-section typically receives only small radiative corrections
of ${\cal O}(\%)$.\footnote{An exception are annihilation processes that are
suppressed at tree-level, but this suppression is lifted at the loop level, see
e.g. Ref~\cite{qq}.} This allows to make robust predictions of the present dark
matter density within a certain model for a given choice of parameters.
Assuming the standard cosmological model, the evolution of the dark matter
density from the time of freeze-out from thermal equilibrium until the present
time can be computed based only on the Boltzmann equation and the thermally
averaged annihilation cross-section (see for example Ref.~\cite{GG}).

Besides predicting a dark matter candidate, most of the Standard Model
extensions also introduce a large spectrum of additional particles. In
supersymmetry, for instance, every Standard Model particle has a supersymmetric
partner. If the mass of one of these particles is close to the WIMP mass, it
would not yet be decoupled from thermal equilibrium at the freeze-out
temperature of the WIMP and thus influences the WIMP annihilation. This
mechanism is called co-annihilation and effectively lowers the present day
relic density \cite{coann}. It is interesting to note that the co-annihilating
particle can have very different quantum numbers than the WIMP, as long as it
decays into a final state which includes the WIMP. In particular, the
co-annihilating particle does not need to be weakly interacting. In some
supersymmetric scenarios, co-annihilation occurs between the lightest
neutralino (as the WIMP) and the stau or stop, which participate in
electromagnetic, and in the latter case also strong, interactions. As a result,
radiative corrections can be very important in co-annihilation processes.

In this letter, QCD corrections to co-annihilation processes are analyzed. For
concreteness, the specific process of neutralino-stop co-annihilation within the
Minimal Supersymmetric Standard Model (MSSM) is considered. 
In many
supersymmetry breaking scenarios, the light stop $\sta$ is predicted to be
the lightest squark state due to large running and mixing effects, so that it is
not improbable that the $\stopone$ can get close in mass to the lightest
neutralino $\neu_1$. This scenario is in particular motivated by electroweak
baryogenesis, see e.g. Ref.~\cite{stop}.
If the mass difference $\mst-\mneu{1}$ is small, and all
other superpartners are significantly heavier than the $\sta$, there are
effectively three contributing processes to the evolution of the dark matter
density,
\begin{equation}
\neu_1\neu_1 \to X, \qquad
\neu_1\sta \to X, \qquad
\sta\sta^{(*)} \to X,
\end{equation}
where $X$ stands for some Standard Model particles. 
The relic density within a stop-neutralino co-annihilation scenario has been first 
calculated in Ref.~\cite{tree}, using tree-level formulae for the these three 
annihilation cross-sections.
For a typical parameter
point with a predominantly bino $\neu_1$ of 118 GeV and a predominantly
R-chiral stop of 138 GeV, the dark matter density is predicted to be $\Omega
h^2 = 0.112$, using {\sc DarkSUSY~4.1} \cite{darksusy}. Here $\neu_1\neu_1$
annihilation contributes only about 5\% to the total averaged annihilation
cross-section, while $\neu_1\sta$ and $\sta\sta^*$ annihilation contribute 85\%
and 10\%, respectively. In the following, QCD radiative corrections to the last
two processes will be studied in detail.

\begin{figure}[tb]
\centering
\raisebox{18mm}{(a)} \raisebox{2.5mm}{\psfig{figure=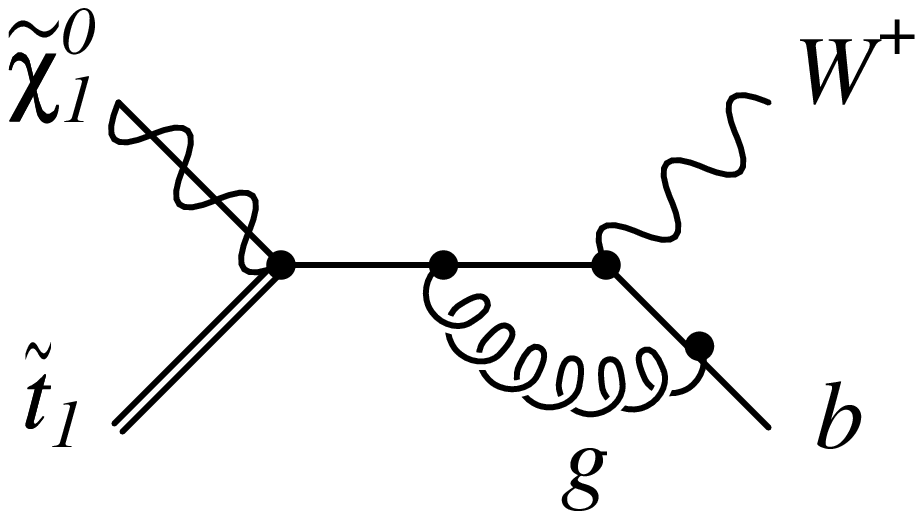, width=5cm}}
\psfig{figure=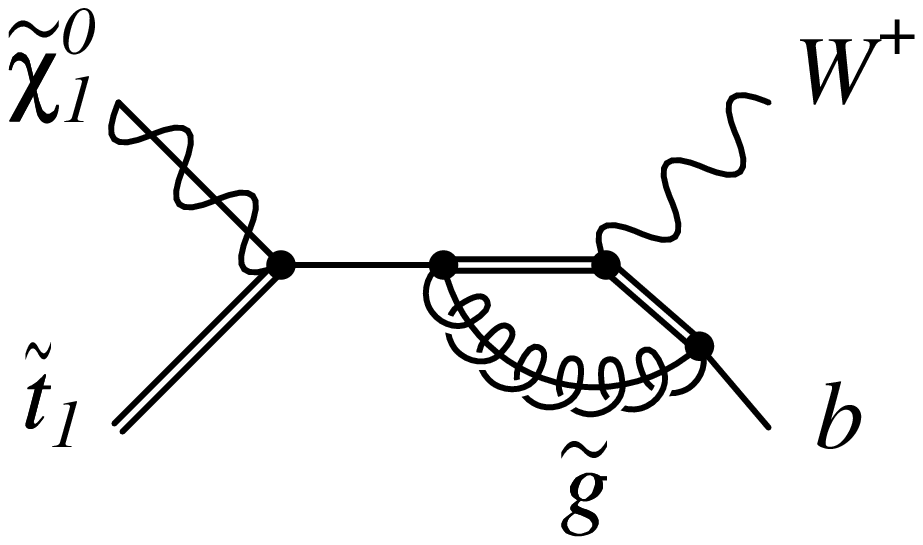, width=5cm}\\
\raisebox{10mm}{(b)} \psfig{figure=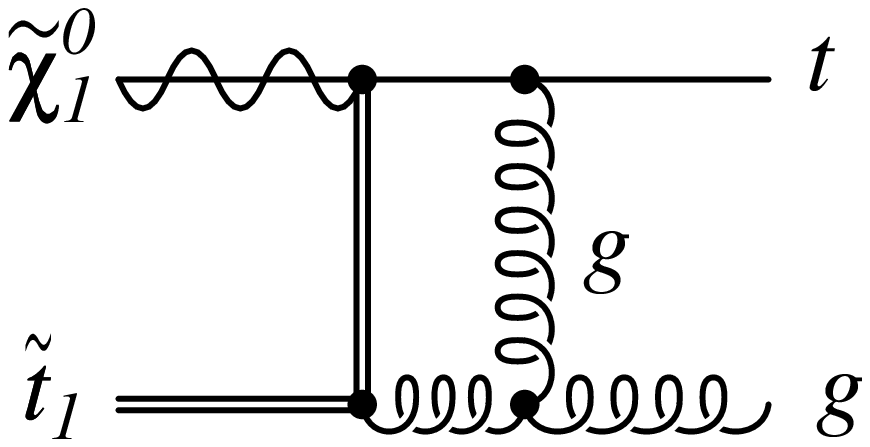, width=5cm}
\psfig{figure=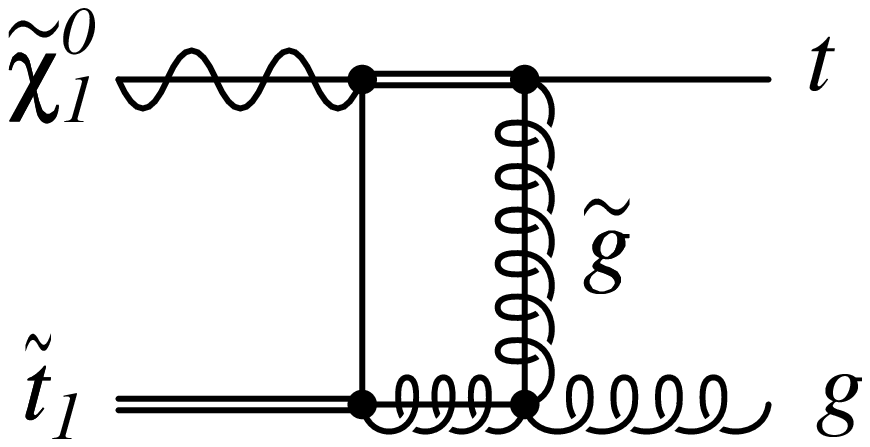, width=5cm}
\mycaption{Some loop diagrams for the ${\cal O}(\as)$ corrections to
(a) $\neu_1\sta \to W^+ b$ and (b) $\neu_1\sta \to t g$.}
\label{fg:dia}
\end{figure}

\paragraph{\boldmath $\neu_1\sta$ annihilation: }
The annihilation of $\neu_1\sta$ into Standard Model particles receives
contributions from several channels. For the aforementioned parameter point with
$\neu_1 \sim \tilde{B}$, $\mneu{1} = 118$ GeV and $\sta \sim \st_{\rm R}$, $\mst
= 138$ GeV, one finds
\begin{align}
\neu_1\sta &\to W^+ b & 49.5\% & \text{ contribution to the thermally averaged
$\neu_1\sta$ cross-section}, \\
\neu_1\sta &\to t g & 47.5\% & , \\
\neu_1\sta &\to t \gamma & 1.0\% & , \\
\neu_1\sta &\to t Z & 2.0\% & ,
\end{align}
while all other kinematically allowed final states are negligible. Since the
first two processes are by far dominant, this work concentrates on the radiative
corrections to $\neu_1\sta \to W^+ b$ and $\neu_1\sta \to t g$. Typical diagrams
for the ${\cal O}(\as)$ virtual corrections are shown in Fig.~\ref{fg:dia}.
%%%%%%%%%%%%%%%%%%%%%%%%%%%%%%%%%%%%%%%%%%%%%%%%%%%%%%%%%%%%%%%%%%%%
%%%%%%%%%%%%%%%%%%%%%%%%%%%%%%%%%%%%%%%%%%%%%%%%%%%%%%%%%%%%%%%%%%%%
Since the tree-level processes include quark-squark-neutralino vertices, the QCD
corrections with gluon exchange cannot be separated from the SUSY-QCD
corrections with gluino exchange. Therefore the complete SUSY-QCD corrections of
order ${\cal O}(\as)$ will be considered.

The technical methods for the calculation are well established. The virtual
loop diagrams have been generated with {\sc FeynArts 3.0} \cite{fa}, while the
algebraic reduction to standard matrix elements was performed with the help of 
{\sc FeynCalc 2.2} \cite{fc}. In the loop integrals UV divergencies occur,
which are canceled through the renormalization of mass, mixing and coupling
parameters. The masses have been renormalized according to the on-shell scheme,
while for the strong coupling constant the $\overline{\text{MS}}$ scheme with
six-flavor SUSY-QCD running has been chosen. Instead of introducing a
counterterm for the stop and sbottom mixing angles \cite{sqmix}, the squark
sector has been renormalized by using matrix-valued mass counterterms
\cite{sqmass}. The two approaches lead to different definitions of the
renormalized mixing angles, but are equivalent (up to higher orders) when
relating physical observables.

\begin{figure}[tb]
\begin{tabular}{p{5.5cm}p{10cm}}
\raisebox{-2.5cm}{\psfig{figure=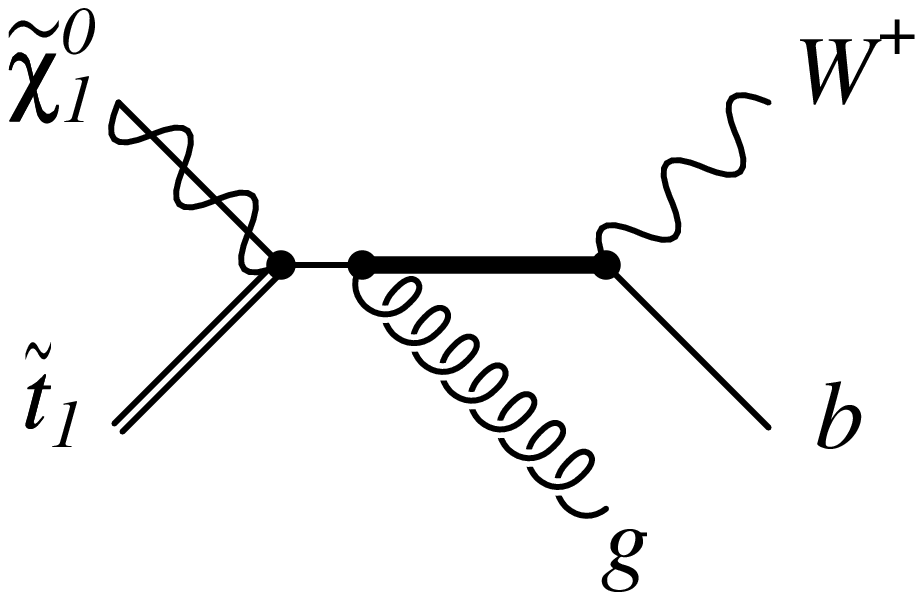, width=5cm}} &
\mycaption{One diagram in the real gluon corrections to $\neu_1\sta \to W^+ b$
where the intermediate top-quark propagator can become on-shell, indicated by a
thick line.}
\label{fg:real}
\end{tabular}
\vspace{-1em}
\end{figure}

The loop diagrams with gluon exchange also lead to the IR divergencies, that
have been regularized by a gluon mass. They need to be combined with the
contributions with real gluon emission in order to arrive at an IR finite
complete result. Here a word of caution is in order for the real corrections to
the process $\neu_1\sta \to W^+ b$. In the diagram shown in Fig.~\ref{fg:real},
%%%%%%%%%%%%%%%%%%%%%%%%%%%%%%%%%%%%%%%%%%%%%%%%%%%%%%%%%%%%%%%%%%%%
%%%%%%%%%%%%%%%%%%%%%%%%%%%%%%%%%%%%%%%%%%%%%%%%%%%%%%%%%%%%%%%%%%%%
the top propagator that is shown as a thick line can be become on-shell. In
this kinematical region, however, it would be a contribution to the process
$\neu_1\sta \to t g$ with the subsequent decay $t \to W^+ b$. Therefore the
terms with a resonant top propagator need to be subtracted from the matrix
element of $\neu_1\sta \to W^+ b + g$. On the other hand, the interference
terms between an amplitude with a resonant top propagator and a non-resonant
amplitude are retained in this channel. 

For the numerical evaluation, the loop integrals are computed with the package
{\sc LoopTools} \cite{lt}, while the phase space for the real corrections is
integrated numerically using Monte Carlo methods. The real radiation
phase space has been
mapped onto the integration variables in an optimized way so as to improve the
result in the regions where the gluon(s) become(s) soft or collinear (the
collinear case only occurs for the process $\neu_1\sta \to t g + g$).
After combining virtual and real corrections, it has been checked that the soft and 
collinear singularities drop out of the total result, so that it does not depend on the gluon mass regulator within phase space integration errors.

\begin{figure}[tb]
\hspace{-7mm}
\epsfig{figure=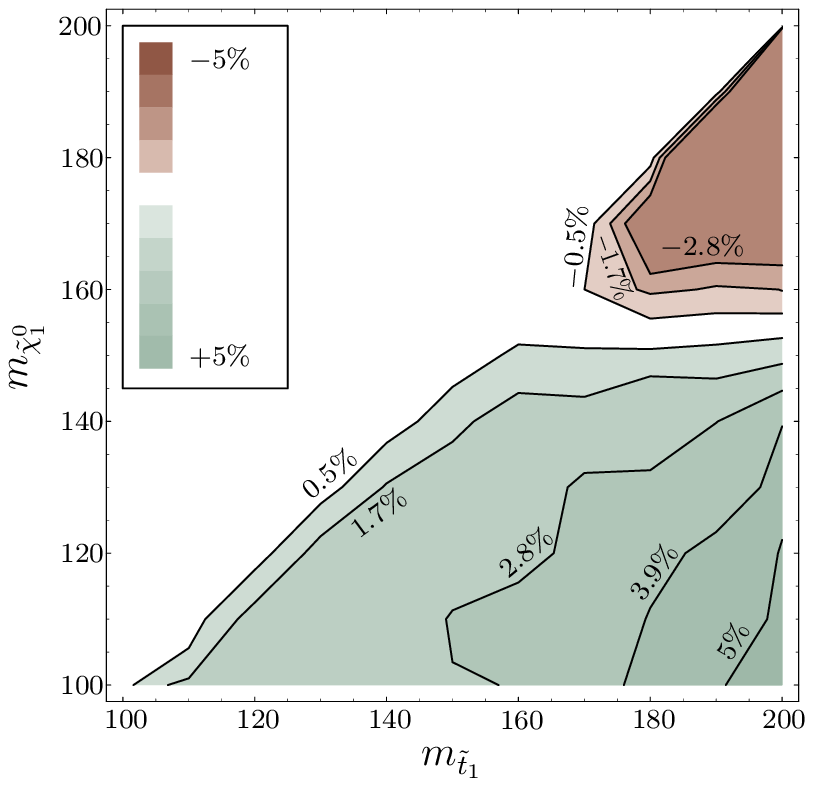, width=8.5cm}
\hspace{-2mm}
\epsfig{figure=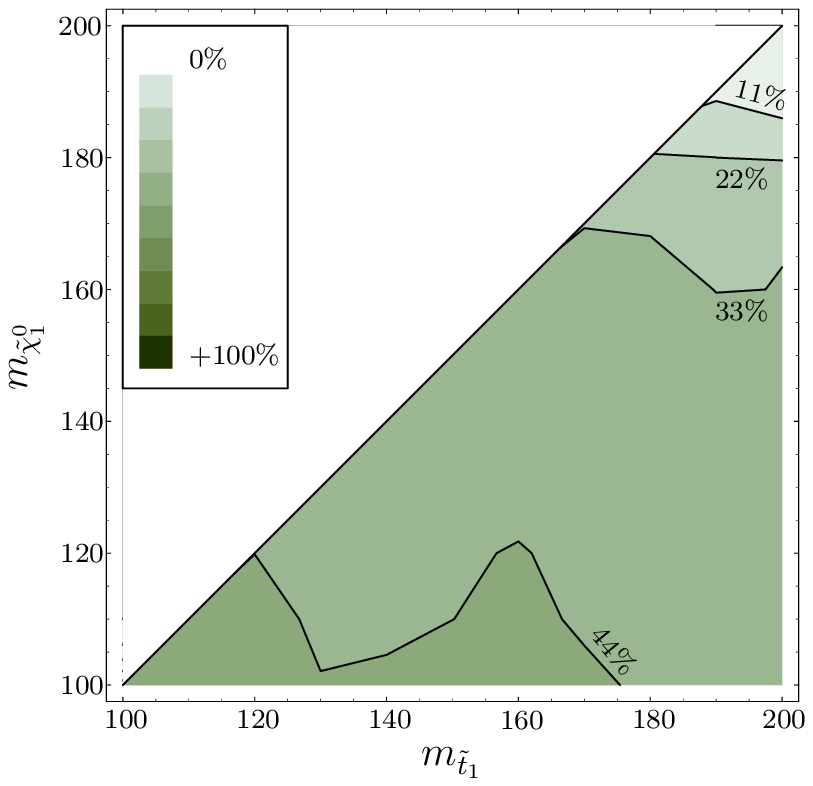, width=8.5cm}
\mycaption{Relative size of the SUSY-QCD corrections to $\neu_1\sta \to W^+ b$
(left) and $\neu_1\sta \to t g$ (right), normalized to the Born cross-sections,
respectively. It is assumed that $\neu_1 = \tilde{B}$ and $\sta = \st_\RR$.
The other relevant supersymmetric parameters are fixed to $m_{\tilde{q}} = 4$
TeV for $\tilde{q} \neq \sta$, $m_{\tilde{g}} = 2$ TeV, and the QCD scale is
set to half the center-of-mass energy, $\mu_0 = \sqrt{s}/2$.}
\label{fg:xt}
\end{figure}

\begin{figure}[tb]
\begin{tabular}{p{9cm}p{7.2cm}}
\raisebox{-4.4cm}{\epsfig{figure=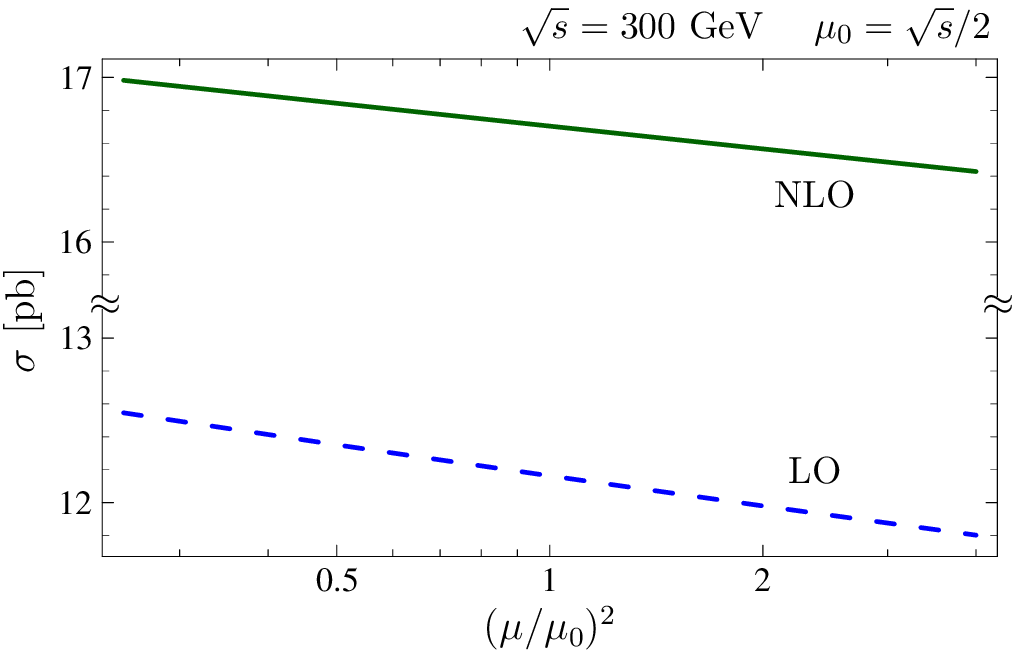, width=9cm}} &
\mycaption{Dependence of the cross-section $\sigma$ for $\neu_1\sta \to t g$ 
on the SUSY-QCD scale $\mu$ at Born (LO) and one-loop (NLO) order. The MSSM
parameters are $\neu_1 = \tilde{B}$, $\sta = \st_\RR$,
$\mneu{1} = 118$ GeV, $\mst = 138$ GeV, $m_{\tilde{q}} = 4$
TeV for $\tilde{q} \neq \sta$, $m_{\tilde{g}} = 2$ TeV.}
\label{fg:scale}
\end{tabular}
\end{figure}

In order to simplify the presentation of the numerical results, one can observe
that the dynamics of co-annihilation are mainly governed by the particle
masses, whereas mixing effects play a minor role. Thus for simplicity, here it
is assumed that the lightest neutralino $\neu_1$ is a pure bino $\tilde{B}$, 
while the
light stop $\sta$ is purely the partner of the right-handed stop\footnote{In addition, a light stop with a sizeable left-chiral component is in conflict with precision measurements of the $Zbb$ vertex.}. The size of
the SUSY-QCD corrections in this scenario is illustrated in Fig.~\ref{fg:xt}.
%%%%%%%%%%%%%%%%%%%%%%%%%%%%%%%%%%%%%%%%%%%%%%%%%%%%%%%%%%%%%%%%%%%%
%%%%%%%%%%%%%%%%%%%%%%%%%%%%%%%%%%%%%%%%%%%%%%%%%%%%%%%%%%%%%%%%%%%%
As can be seen from the figure, the corrections to  $\neu_1\sta \to W^+ b$
amount to only a few per-cent, while the process $\neu_1\sta \to t g$ receives
large positive corrections that can reach roughly 50\%. Note that the size of
the corrections is much larger than the scale uncertainty of the Born
cross-section, see Fig.~\ref{fg:scale}.
%%%%%%%%%%%%%%%%%%%%%%%%%%%%%%%%%%%%%%%%%%%%%%%%%%%%%%%%%%%%%%%%%%%%
%%%%%%%%%%%%%%%%%%%%%%%%%%%%%%%%%%%%%%%%%%%%%%%%%%%%%%%%%%%%%%%%%%%%
For the next-to-leading order result, the scale uncertainty is only slightly
reduced with respect to the leading-order cross-section.

\paragraph{\boldmath $\sta\sta^*$ annihilation: }
As mentioned above, for small stop masses $\sta\sta^*$ 
annihilation typically contributes much less to
the total thermally averaged co-annihilation cross-section then $\neu_1\sta$
annihilation. Nevertheless, there are potentially very large Coulombic gluon
corrections that can be important for the computation of the relic density.
These corrections arise from the exchange of gluons between the stop and
anti-stop. Since during the phase of freeze-out, the stops and anti-stops are
slowly moving ($E_{{\rm kin},\st} \approx T_{\rm freeze-out} \ll \mst$), long-range
gluon exchange effects lead to a strong enhancement of the $\sta\sta^*$
annihilation cross-section. A similar enhancement effect due to QED corrections
in $\neu_1$-$\cha^\pm_1$ co-annihilation in the focus point region has been
studied in Ref.~\cite{hisano}.

%%%%%%%%%%%%%%%%%%%%%%%%%%%%%%%%%%%%%%%%%%%%%%%%%%%%%%%%%%%%%%%%%%%%
\begin{figure}[tb]
\vspace{-5mm}
\begin{tabular}{p{4cm}p{6cm}p{5.3cm}} 
\raisebox{-5mm}{(a)}\hspace{-5mm}\raisebox{-22mm}{\psfig{figure=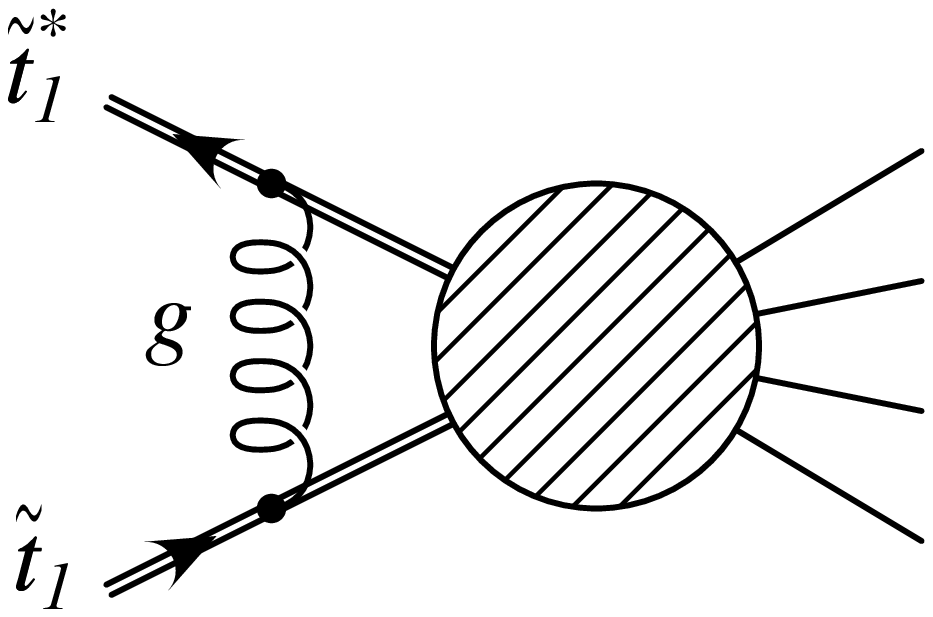, width=3.6cm}} &
\raisebox{-5mm}{(b)}\hspace{-5mm}\raisebox{-20mm}{\psfig{figure=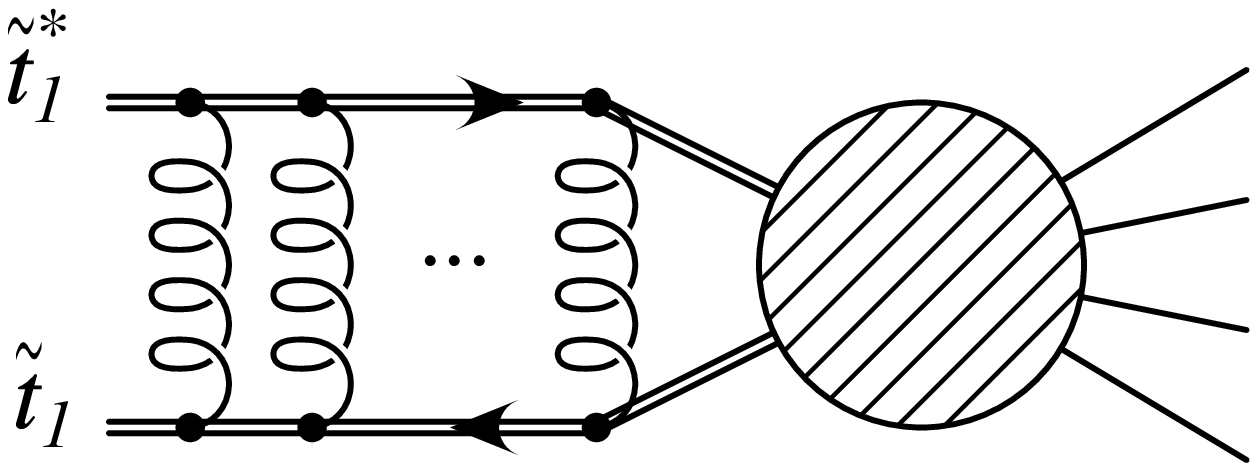, width=4.8cm}} &
\mycaption{Diagrams for the (a) leading and (b) resummed Coulombic gluon
corrections to $\sta\sta^*$ annihilation.}
\label{fg:coul}
\end{tabular}
\vspace{-1em}
\end{figure}
%%%%%%%%%%%%%%%%%%%%%%%%%%%%%%%%%%%%%%%%%%%%%%%%%%%%%%%%%%%%%%%%%%%%
The leading contribution shown in Fig.~\ref{fg:coul}~(a)
leads to a correction
\begin{equation}
\sigma_{\rm coul} = \frac{2\as\pi}{3 v} \sigma_{\rm Born},
\end{equation}
where $v$ is the relative velocity between the $\sta$ and $\sta^*$. The
divergence for $v \to 0$ is the well-known Coulomb singularity. In the case of
dark matter annihilation, the Coulomb singularity is naturally cut off by the
finite temperature in the early universe. Nevertheless, $v$ is typically of the
same order as $\as$, so that higher-order effects need to be included.
The exchange of
$n$ gluons, as in Fig.~\ref{fg:coul}~(b), generates a correction factor
$\propto (\as/v)^n$. These effects can be systematically resummed within
non-relativistic QCD (NRQCD), which is an effective theory of full QCD valid at
low velocities. Within NRQCD, the problem is described by the Schr\"odinger
equation with the Coulombic QCD potential $V(\mbox{\boldmath $r$})$
\cite{schreq},
\begin{equation}
\left[-\frac{\Delta}{\mst} + V(\mbox{\boldmath $r$})\right] \Psi(\mbox{\boldmath
 $r$})
= (E + i \Gamma)\, \Psi(\mbox{\boldmath $r$}), \qquad
V(\mbox{\boldmath $r$}) = - C_{\rm F} \frac{\alpha_{\rm s}}{r},
\label{eq:schreq}
\end{equation}
where $C_{\rm F}$ is the Casimir operator of the SU(3) fundamental
representation, and
$E$ and $\Gamma$ are the energy and decay width of the $\langle \sta\sta^*
\rangle$ system. The correction factor to the cross-section is then given by
\begin{equation}
\frac{\sigma_{\rm coul,resum}}{\sigma_{\rm Born}} = |\Psi(0)|^2.
\end{equation}
The QCD Coulomb potential receives important higher order corrections from
gluon self-interaction and fermion loop effects \cite{clnlo,clnnlo}. They can be
calculated most easily in momentum space and are given by
\begin{equation}
V(q^2) = - C_F \frac{4\pi\alpha_{\rm s}}{q^2}
\left[ 1+ \frac{\alpha_{\rm s}}{4\pi}\left(\frac{31}{9} C_A -\frac{20}{9} T_F n_
f +
  \beta_0 \, \log (\mu^2/q^2) \right) + {\cal O}(\as^2) \right].
\end{equation}
For a typical QCD scale choice $\mu \sim \mst$, the logarithm $\log (\mu^2/q^2)
= \log (1/v)$ can become large, and it is advantageous to resum contributions
of order ${\cal O}(\as\log^n v)$ in a similar way as the ${\cal O}((\as/v)^n)$
terms. This can be achieved by more elaborate effective theory frameworks, such
as velocity non-relativistic QCD (vNRQCD) \cite{vnrqcd}  or potential
non-relativistic QCD (pNRQCD) \cite{pnrqcd}. In this work, however, only a
simple first estimate of the threshold corrections shall be obtained, for which
the framework of NRQCD as in eq.~\eqref{eq:schreq} is sufficient.

The leading contribution to the $\sta\sta^*$ annihilation cross-section arises
from the lowest S-wave state, i.e. the $\langle \sta\sta^*
\rangle_{1S}$ state. In eq.~\eqref{eq:schreq}
the decay width $\Gamma$ of the stopponium state $\langle \sta\sta^*
\rangle_{1S}$ needs to be included. Taking into account the two largest decay
channels, $\langle \sta\sta^*
\rangle_{1S} \to gg, \; W^+W^-$, the partial widths at Born level read
\begin{equation}
\Gamma[\langle \sta\sta^*\rangle_{1S} \to gg] = \frac{448 \, \as^5 \,\mst}{243},
\qquad
\Gamma[\langle \sta\sta^*\rangle_{1S} \to WW] =
\frac{\as^3 y_{\rm t}^4 \, \mst}{6\pi^2} 
\biggl[ 1 - \frac{\MW^2}{\mst^2} + \frac{3 \MW^4}{4 \mst^4} \biggr],
\end{equation}
where the second formula is valid in the limits $M^2_{h^0} \ll 4 \mst^2 \ll
M^2_{A^0}$ and $\tan\beta \gg 1$. $y_{\rm t}$ is the top Yukawa coupling. For
$\mst = 138$ GeV, the total width is $\Gamma \approx 5$ MeV, and thus almost
negligible. The stop $\sta$ also has an intrinsic width, with for small mass
differences $\mneu{1}-\mst$ is however smaller than 1 MeV \cite{stdecay} and
therefore can be safely ignored.

\begin{figure}[tb]
\begin{tabular}{p{9cm}p{6.7cm}}
\raisebox{-4.4cm}{\epsfig{figure=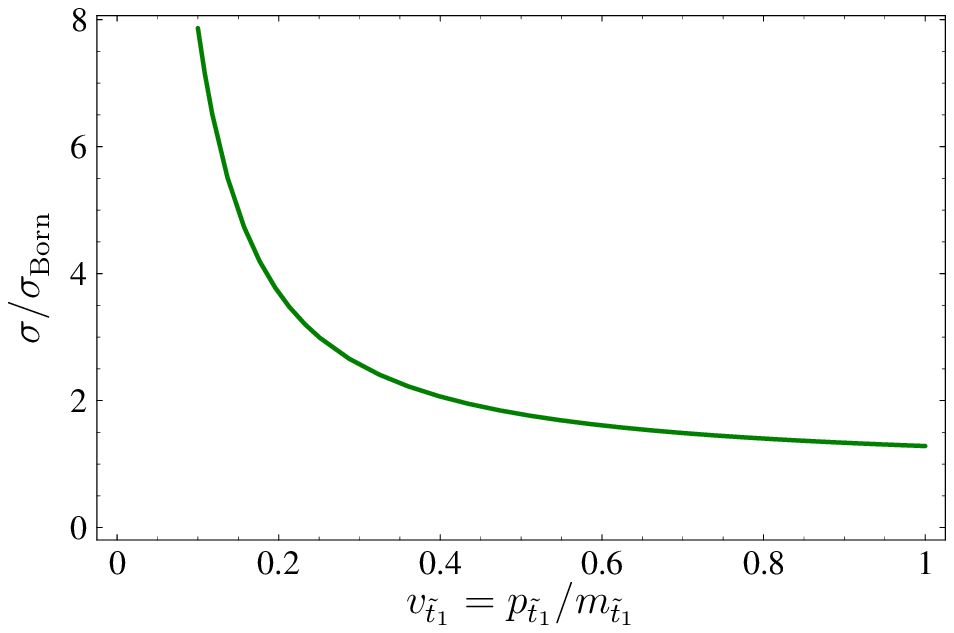, width=9cm}} &
\mycaption{Cross-section for $\sta\sta^*$ annihilation including Coulombic
QCD corrections near threshold relative to the Born cross-section, as a function
of the velocity $v=(1- 4 \mst^2/s)^{1/2}$. The QCD scale has been fixed to $\mu =
122$ GeV, using five-flavor running of $\as$ in normal QCD (not SUSY-QCD).}
\label{fg:coul2}
\end{tabular}
\vspace{-1em}
\end{figure}

For the numerical evaluation, the Schr\"odinger equation with the QCD potential
at two-loop order \cite{clnnlo} is solved numerically. The result is shown in
Fig.~\ref{fg:coul2}, which illustrates that for $v \lesim 0.4$ the Coulombic
correction can be larger than 100\% of the tree-level cross-section.
%%%%%%%%%%%%%%%%%%%%%%%%%%%%%%%%%%%%%%%%%%%%%%%%%%%%%%%%%%%%%%%%%%%%
%%%%%%%%%%%%%%%%%%%%%%%%%%%%%%%%%%%%%%%%%%%%%%%%%%%%%%%%%%%%%%%%%%%%
For very low values of $v$, the results of this calculation are not reliable,
since the effect of $\sta\sta^*$ bounds states needs to be
taken into account properly. However, as already mentioned above, the formation
of bound states is inhibited by the non-zero temperature during freeze-out, so
that the region $\mst v \ll T_{\rm freeze-out}$ is not relevant for stop
annihilation. In other words, in the thermally averaged integration over the
annihilation cross-section, the region around $v=0$ has vanishing integration
measure, so that no special treatment for the Coulomb singularity is necessary.

\begin{figure}[tb]
\hspace{-7mm}
\epsfig{figure=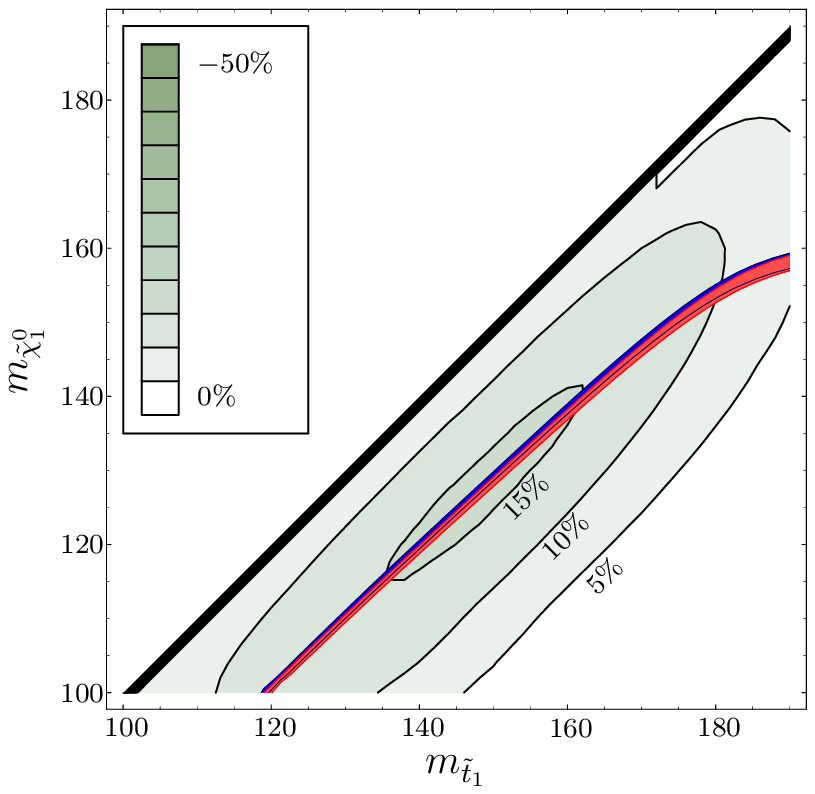, width=8.5cm}
\hspace{-2mm}
\epsfig{figure=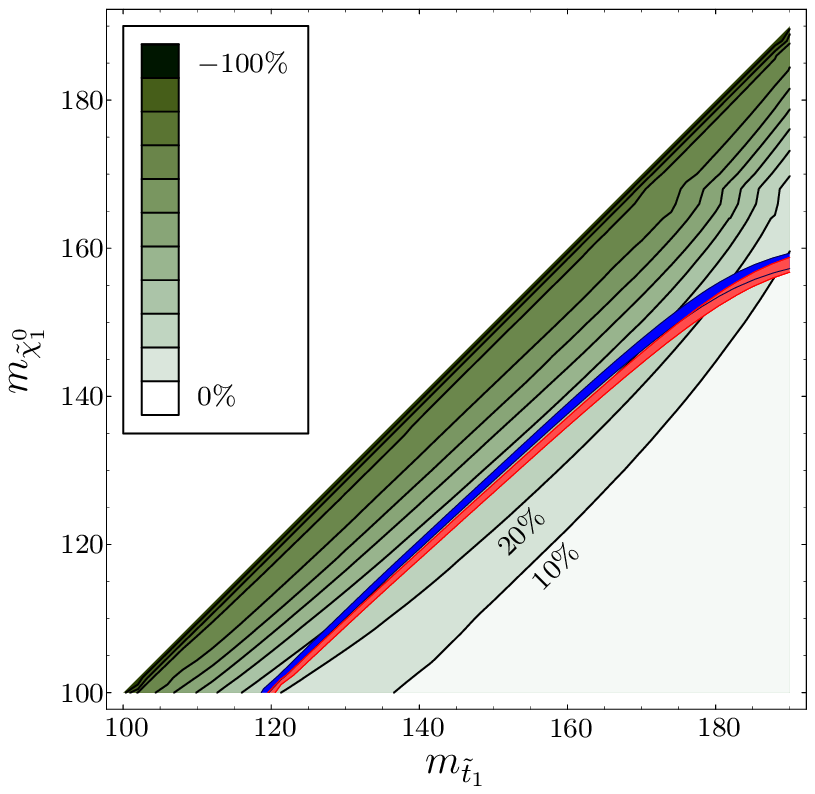, width=8.5cm}
\mycaption{Relative shift of the predicted dark matter density $\Omega_{\rm
CDM}$ due to radiative corrections.
{\rm Left:} Only SUSY-QCD corrections to $\neu_1\sta$ annihilation.
{\rm Right:} SUSY-QCD corrections to $\neu_1\sta$ annihilation \emph{and} 
Coulombic QCD to $\sta\sta^*$ annihilation.
The blue/dark and red/light bands indicate the parameter region in agreement with the
observed value of $\Omega_{\rm
CDM}$ without and with the radiative corrections, respectively.
The parameters have been chosen as in Fig.~\ref{fg:xt}.}
\label{fg:res}
\end{figure}

\paragraph{Effect on relic density:}
In order to study the effect of the radiative corrections to $\neu_1\sta$ and
$\sta\sta^*$ annihilation, as presented in the previous sections, 
they have been implemented into {\sc DarkSUSY~4.1} \cite{darksusy}. The effect
of the corrections for different neutralino and stop masses is shown in
Fig.~\ref{fg:res}.
%%%%%%%%%%%%%%%%%%%%%%%%%%%%%%%%%%%%%%%%%%%%%%%%%%%%%%%%%%%%%%%%%%%%
%%%%%%%%%%%%%%%%%%%%%%%%%%%%%%%%%%%%%%%%%%%%%%%%%%%%%%%%%%%%%%%%%%%%
In the left panel only the SUSY-QCD corrections to $\neu_1\sta$ annihilation are
incorporated, which can change the predicted relic density by up to about 15\%.
This is comparable with the current uncertainty of $\Omega_{\rm
CDM}$ from astrophysical observations \cite{cos}. In the right panel, in
addition the Coulombic QCD corrections to $\sta\sta^*$ annihilation have been
taken into account. It can be seen that for small $\sta$-$\neu_1$ mass
differences, the predicted value of $\Omega_{\rm CDM}$ can shift by more than
50\% due to these effects. Even in the region which is an agreement with the
current value $\Omega_{\rm CDM} h^2 = 0.1106^{+0.0056}_{-0.0075}$, the loop
correction effects are larger than the current error from Ref.~\cite{cos}. This
is indicated by the blue band in the figure, which is the astrophysically
allowed region for tree-level annihilation cross-sections, 
and the red band, which is the same after
including the radiative corrections.

\vspace{2ex}

In summary, the leading ${\cal O}(\as)$ corrections to neutralino-stop
co-annihilation have been calculated. The process is typical for any model and
scenario, where a WIMP dark matter particle annihilates together with a
strongly interacting particle that can decay into the WIMP. It was found that
the $\neu_1\sta$ annihilation process receives sizeable radiative corrections
that modify the predicted dark matter relic density by 5--15\%. On the other hand, the
$\sta\sta^*$ annihilation process receives very large corrections, that are
associated with the Coulomb singularity near threshold. The effect of these
Coulombic QCD corrections can have an impact on the computed relic density by
more than 50\% for small stop-neutralino mass differences. In a previous work by
Hisano {\it et al.} \cite{hisano}, similarly large
effects have been found as a result of QED threshold corrections in
neutralino-chargino co-annihilation.

Since the leading radiative corrections presented in this paper turn out to be
large, the results are expected to still have a sizeable theoretical error.
For a reliable prediction of the dark matter
relic density, the theoretical calculations need to be refined beyond the
techniques employed in this work, including the resummation of large logarithms.
However, it is  interesting to note that the effect of the large QCD
threshold corrections to $\sta\sta^*$ annihilation can also be derived from a
measurement of the pair production cross-section for $e^+e^- \to \sta\sta^*$ at
a future linear collider, see e.g. Ref.~\cite{stopsLC}.

\paragraph{Acknowledgments:}
The author is grateful to P.~Gondolo for helpful discussions and to S.~Kraml and
P.~Gondolo for careful reading of the manuscript.

%%%%%%%%%%%%%%%%%%%%%%%%%%%%%%%%%%%%%%%%%%%%%%%%%%%%%%%%%%%%%%%%

\end{document}